\documentclass[aps,showpacs,prl,twocolumn]{revtex4}
\usepackage{graphics}

\begin{document}

\preprint{{DOE/ER/40762-277}\cr{UMPP\#03-042}}


\count255=\time\divide\count255 by 60
\xdef\hourmin{\number\count255}
  \multiply\count255 by-60\advance\count255 by\time
 \xdef\hourmin{\hourmin:\ifnum\count255<10 0\fi\the\count255}

\newcommand{\xbf}[1]{\mbox{\boldmath $ #1 $}}

\title{Nucleon-Nucleon Scattering under Spin-Isospin Reversal in Large-$N_c$ QCD}

\author{Thomas D. Cohen}
\email{cohen@physics.umd.edu}

\affiliation{Department of Physics, University of Maryland,
College Park, MD 20742-4111}

\author{Daniel C. Dakin}
\email{dcdakin@physics.umd.edu}

\affiliation{Department of Physics, University of Maryland,
College Park, MD 20742-4111}

\date{March, 2003}

\begin{abstract}
The spin-flavor structure of certain nucleon-nucleon scattering
observables derived from the large $N_c$ limit of QCD in the
kinematical regime where time-dependent mean-field theory is
valid is discussed. In previous work \cite{TB} this regime was
taken to be where the external momentum was of order $N_c$ which
precluded the study of differential cross sections in elastic
scattering. Here it is shown that the regime extends down to
order $N_c^{1/2}$ which includes the higher end of the elastic
regime. The prediction is that in the large $N_c$ limit,
observables describable via mean-field theory are unchanged when
the spin and isospin of either nucleon are both flipped. This
prediction is tested for proton-proton and neutron-proton elastic
scattering data and found to fail badly. We argue that this
failure can be traced to a lack of a clear separation of scales
between momentum of order $N_c^{1/2}$ and $N_c^1$ when $N_c$ is
as small as three. The situation is compounded by an anomalously
low particle production threshold due to approximate chiral
symmetry.

\end{abstract}


\maketitle

Calculating observables for nucleon-nucleon scattering directly
from QCD is a formidable task. Given our experience with other
hadronic observables, one might hope to make qualitative
predictions by exploiting the large $N_c$ limit of QCD; the limit
in which the number of colors, $N_c$, becomes large and $1/N_c$
is regarded as an expansion parameter~\cite{TH,EW}. Experience in
the single baryon sector shows that one of the striking features
of large $N_c$ QCD is a simple spin-flavor structure associated
with a contracted $SU(2 N_f)$ symmetry which becomes exact as
$N_c \rightarrow \infty$ \cite{C1}. There has been considerable
work over the past several years on the spin-flavor structure of
the nucleon-nucleon potential appropriate for low momentum ($p
\sim N_c^0$) scattering. In particular, it has been established
that the leading order terms in the potential are those for which
the $t$-channel exchange has the same angular momentum and
isospin \cite{C2}. Furthermore, this picture is consistent with
both single-meson and multiple-meson exchange \cite{C3}. While
this is of interest theoretically, it is of limited direct
phenomenological consequence since the potential cannot be
directly measured. Recently, ref.~\cite{TB} noted that the
spin-isospin dependence of certain nucleon-nucleon scattering
observables was calculable up to $1/N_c$ corrections. In
particular, they showed that for large $N_c$ those observables
which are calculable in principle from time-dependent mean-field
theory (TDMFT) are invariant under simultaneous spin and isospin
flips of either incident particle.

Before proceeding, it is necessary to clarify exactly what this
restriction to observables calculable in TDMFT implies.  Firstly,
as originally noted by Witten~\cite{EW} there is a kinematic
constraint since TDMFT is inapplicable for momenta of order
$N_c^0$, but becomes exact at large $N_c$ for momenta of order
$N_c^1$ (which implies velocities of order $N_c^0$ since the
nucleon mass goes $N_c^1$).  Secondly, mean-field theory only
becomes exact in the large $N_c$ limit for observables without
multi-particle correlations, {\it i.e.} for observables
associated with the collective flow of currents such as baryon
number or energy. Examples of such observables include the
differential cross section for baryon number (in the form of
nucleons) to flow out from a collision (which in general produces
many mesons) at a fixed angle.  Since the observable integrates
over all of the final meson states, it is independent of multiple
particle correlation and depends only on the collective flow of
baryon number. Thus it is calculable in TDMFT and subject to the
spin-isospin constraint given above.

In ref.~\cite{TB} it was pointed out that the most familiar
nucleon-nucleon scattering observables--the total cross section,
the total-elastic cross section and the differential cross section
for elastic scattering--were not in the class of observables
describable in mean-field theory in the kinematic regime of $p
\sim N_c^1$.  The total cross section includes contributions from
nearly-forward scattering and forward scattering is not accessible
in semi-classical treatments such as TDMFT~\cite{LL}. Similarly,
the total-elastic cross section is inaccessible.  The differential
elastic cross section in this kinematic regime is inaccessible in
mean-field theory for a different reason. When $p \sim N_c$, one
is well above the threshold for meson production (which scales as
$p \sim N_c^{1/2}$ for an incident kinetic energy $p^2/(2 M_N)
\sim N_c^0$) and hence the typical collision is inelastic.  The
extraction of the elastic differential cross section thus depends
strongly on multi-particle correlations in that it totally
excludes the many-particle final states.  As such it cannot be
computed in TDMFT and the spin-isospin constraint does not apply.

As a practical matter, the exclusion of these more standard
observables has made it very difficult to test the large $N_c$
spin-flavor predictions of ref.~\cite{TB}.  We know of no
nucleon-nucleon scattering experiments in the kinematic regime of
$p \sim N_c$ (which we can take to be $p \sim M_N$) where the
data has been reported as a differential cross section for the
outgoing nucleons with the outgoing mesons integrated over, and
for which data is available for both protons and neutrons with
polarized targets and/or beams.

Clearly it would be very useful to have some experimentally
accessible observable that is also computable in TDMFT.  Ideally,
we would hope to find some observable which is readily available
experimentally and of the correct class.  Since the most readily
available data is for the total and elastic cross sections
discussed above, it would be extremely useful if there were some
circumstance in which one of these may be reliably computed in
TDMFT. The argument that forward scattering is not computable in
TDMFT is quite generic and holds for all regimes. In contrast,
the argument that the differential cross section for elastic
scattering is not computable in TDMFT depended on Witten's
kinematic condition that $p\sim N_c$ for TDMFT to be valid.  The
key theoretical point of the present paper is that Witten's
condition is overly restrictive.  While TDMFT is clearly
inapplicable in the regime where $p \sim N_c^0$ and is clearly
applicable in the regime of $p \sim N_c^1$, there is an
intermediate regime of $p \sim N_c^{1/2}$.  This intermediate
regime includes the threshold for meson production so that one
can be just below the meson production threshold and still be in
this regime. Here we will argue that TDMFT is, in fact,
applicable in the intermediate regime at sufficiently large $N_c$
and as such one can have observables associated with current
flows. Below meson production, however, the differential cross
section for baryon flow into a given direction is simply the
elastic differential cross section.

We begin by reviewing the argument that led to the spin-isospin
constraint under investigation. First, let us consider nucleons
in the large $N_c$ limit. Witten argued that they could be
described by a Hartree-type approximation. From this, it was
deduced that the nucleon size is of order unity and that the mass
is of order $N_c$ \cite{EW}. Witten also noted that quantities in
large $N_c$ QCD scale with $1/N_c$ in the same manner that the
analogous quantities in soliton models scale with $g^2$, where $g$
is the coupling constant. This suggests that nucleons can be
interpreted as solitons of mesonic fields. This notion was
exemplified in terms of the Skyrme model---a model in which
baryons are topological solitons of a nonlinear chiral
field~$U_0$. The large $N_c$ limit corresponds to the classical
solution \cite{C4}. For the ground state baryons the field takes
the {\it hedgehog} form
$U_0(x)=\exp[i\theta(r)\vec{\tau}\cdot\hat{r}]$, where the
isospin and coordinate space vectors are coupled; the detailed
functional form of the chiral angle, $\theta(r)$, is irrelevant
for our purpose. The hedgehog breaks both angular momentum and
isospin and thus corresponds to states of more than one spin and
isospin. To get physical states with well-defined spin and isospin
we need to project the hedgehog, which at large $N_c$ can be done
via a semi-classical ``cranking'' technique. In this formalism a
set of collective coordinates associated with isospin rotations is
introduced $\{a_i, a_{0}^2+\vec{a}\cdot\vec{a}=1\}_{i=0}^3$  and
it is convenient to represent them as a matrix:
$A=a_0+\vec{a}\cdot\vec{\tau}\in SU(2)$. Then the matrix
$U(t)\equiv AU_{0}A^{\dagger}$ contains all the information about
the collective dynamics. These dynamics can be quantized
~\cite{ANW,CB} to yield states of definite spin ($J$) and isospin
($I$), described by wavefunctions proportional to the Wigner $D$
functions:~$\Psi(A)\propto D_{m,m_I}^{J=I}(A)$.

Now, let us now consider nucleon-nucleon scattering in light of
the large $N_c$ limit in the context of the Skyrme model.  We use
the Skyrme model here to make the argument concrete, but the
result is expected to be completely model independent.  As noted
by Witten, the strength of the nucleon-nucleon interaction is of
order $N_c$ and the scattering process can be described by TDMFT.
Since the interaction and mass is of order $N_c$, the nucleon's
momentum must also be of order $N_c$ ($p\sim N_c$) to ensure a
smooth limit for scattering observables as
$N_c\!\rightarrow\!\infty$. Initially, the nucleons are well
separated and approach each other along the beam direction,
$\hat{n}$ with an impact parameter, $\hat{b}$. In the Skyrme
picture each is taken as a hedgehog, parameterized by its own $A$
matrix and wavefunction $\Psi(A)$, with spin ($m^{(1,2)}$) and
isospin ($m_{I}^{(1,2)}$) projections. In TDMFT, the observable
must be written in terms of an operator $f(A_1,A_2)$. After
quantization, this corresponds to integrating over the collective
variables weighted by the appropriate collective wavefunctions.
The resulting operator loses any explicit dependence on the
collective variables and instead are expressed in terms of the
spin or isospin. Moreover, these quantities only appear in the
form $\vec{S}/{\cal I}$ or $\vec{I}/{\cal I}$, where $\cal I$ is
the hedgehog Skyrmion's moment of inertia. Since $\cal I$ is of
the order $N_c$~\cite{ANW}, these terms may be neglected at
leading order. To compute an observable, we must calculate the
expectation value of the associated operator and integrate over
the impact parameter and collective variables . Reference
\cite{TB} implemented this procedure and arrived at the following
form for an observable:
\begin{eqnarray}
\sigma_{m^{(1)},m_{I}^{(1)},m^{(2)},m_{I}^{(2)}}^O \hspace{1.25in}
\nonumber
\\ =\int d^{2}\vec{b}\:dA_1\:
dA_2\left|D_{m^{(1)},m_{I}^{(1)}}^{1/2}(A_1)\right|^{2}\nonumber
\\
\times \left|D_{m^{(2)},m_{I}^{(2)}}^{1/2}(A_2)\right|^{2}
f(A_1,A_2)\,,\label{eq1}
\end{eqnarray}
where $\int dA$ is the invariant Haar measure on $SU(2)$. We note
that the spin and isospin projection dependence is contained in
the Wigner $D$ functions and that the time reversal invariance of
the rotation operator dictates that $\left|D_{m,
n}^{1/2}(A)\right|^{2}=\left|D_{-m, -n}^{1/2}(A)\right|^{2}$. This
finally leads us to the condition that observables are unchanged
at leading order in $1/N_c$ when one simultaneously flips both
the spin and isospin of either nucleon. As an example, consider
this condition for an observable which averages over the
direction of all detected particles, thus leaving the beam
direction, $\hat n$, and spins as the only remaining vectors.
Insisting that it be invariant under time reversal and parity, we
arrive at the general form
\begin{eqnarray}
\sigma^O&=&A_0+B_I(\vec{\sigma}^{(1)}\cdot\vec{\sigma}^{(2)})(\vec{\tau}\,^{(1)}\cdot\vec{\tau}\,^{(2)})\nonumber \\
        & &+C_I(\hat n\cdot\vec{\sigma}^{(1)})(\hat
n\cdot\vec{\sigma}^{(2)})(\vec{\tau}\,^{(1)}\cdot\vec{\tau}\,^{(2)}),\label{eq2}
\end{eqnarray}
where $A_0, B_I, C_I$ are functions of energy. If one averages
over the spins, as we would for an unpolarized measurement, we
are left with only the isoscalar $A_0$. Therefore, the result
will be independent of isospin.

The key issue which must be addressed is the kinematic conditions
under which the preceding analysis is valid. Clearly we must
assume that $N_c$ is sufficiently large. Moreover, Witten noted
that at low momenta, $p \sim N_c^0$, one does not have a smooth
large $N_c$ limit for scattering observables, while at larger
momenta, $p \sim N_c$, those observables calculable in TDMFT will
have smooth limits provided we keep the velocities fixed as we
take $N_c$ to infinity. The issue of relevance here is the
determination of the minimum velocity for TDMFT to be valid. To
focus sharply on this issue consider some quantity, $\Theta$,
which is computable in TDMFT. At large $N_c$, $\Theta$ is a
function of the velocity (which scales as $p/N_c$). Consider now
the limit of this function as the velocity approaches zero. The
quantity involves a double limit, $N_c \rightarrow \infty$ and $v
\rightarrow 0$, and the limit is not uniform. If the large $N_c$
limit is taken first, one can smoothly take $v$ down to zero.
However, if one takes $v \rightarrow 0$ first, then for any given
$N_c$ one passes through the regime $p \sim N_c^0$ prior to
reaching $v=0$. However, the regime $p \sim N_c^0$ is outside the
validity of TDMFT, and one expects $\Theta$ to have no smooth
large $N_c$ limit.

Consider the combined zero velocity and large $N_c$ limits taken
in the following way:
\begin{equation}
N_c^{-1} \rightarrow 0 \;\;,\; v \rightarrow 0 \; \;\; {\rm with}
\;\;\; v N_c^{1/2} \sim p N_c^{-1/2} \rightarrow {\rm fixed} \;.
\label{condition}
\end{equation}
This limit is interesting in that it corresponds simultaneously
to zero velocity and infinite momentum. As we shall argue below,
the infinite momentum nature of the limit justifies the use of
TDMFT and, thus, this combined limit is both smooth and has the
same result as taking the large $N_c$ limit first and then the
zero velocity limit.

Formally, we see that as $N_c \rightarrow \infty$, velocities of
order $N_c^{-1/2}$ are in the semi-classical regime and, hence,
can be described by TDMFT.  Moreover, the usual differential cross
section in this regime and just below the meson production
threshold is exactly the same as the observable associated with
the differential flow of baryon number through various angles and
as such is computable in TDMFT.

The argument that the limiting procedure given above leads to a
quantity computable in TDMFT is straightforward. TDMFT is
semi-classical in nature and the fundamental issue is under what
kinematic conditions the classical behavior sets in. We know that
the momentum must be high enough so that the nucleon's deBroglie
wavelength is small compared to the size of the interaction
region. Also, the variation in the effective wavelength as it
traverses the interaction region must remain small compared to
the wavelength itself. This condition must hold even when the
interaction strength becomes large, as in our case since it
scales as $N_c^1$. Simple potential model calculations with
masses and potential strengths both scaling as $N_c^1$ show that
the semi-classical condition is reached for $p \sim N_c^{1/2}$.
With this it is highly plausible that the semi-classical
approximation is valid for $p \sim N_c^{1/2}$ for large $N_c$ QCD.
Since the key issues here are spin and flavor, one condition for
the semi-classical regime to be valid is that the nucleon's
velocity must be rapid enough so that the characteristic rotation
time of a hedgehog is long compared to the characteristic
interaction time.  The characteristic rotation time depends on
the moment of inertia ${\cal I}$ and is given by ${\cal I}/S$
which is order $N_c$; the characteristic interaction time is
$R/v$, where $R \sim N_c^0$ is the size of the nucleon.
Reference~\cite{TB} noted that the condition holds for velocities
of order $N_c^0$ which yields interaction times of order unity.
Here we simply note that the condition also holds in the large
$N_c$ limit for $v \sim N_c^{-1/2}$ which yields interaction times
of order $N_c^{1/2}$ which is parametrically much smaller than
the rotation times.

We also note that the semi-classical approach is known to fail for
forward scattering~\cite{LL}, meaning one cannot make valid
predictions in the small angular region around the beam direction,
$\hat{n}$.

Typical elastic nucleon-nucleon scattering experiments consist of
a proton beam with specified polarization incident on a proton or
neutron target. Detectors in the reaction chamber can record the
lab angle of scattering, the energy, and the polarization of the
scattered nucleon. From this information, various observables can
be calculated to describe the event. Let us consider two
examples: the unpolarized differential cross section
($d\sigma/d\Omega$) for elastic collisions and the polarization
asymmetry in the cross section ($\cal A$).

The unpolarized differential cross section, $d\sigma/d\Omega$, is
obtained by counting the number of nucleons scattered into the
solid angle $d\Omega$ and averaging over the polarization of the
beam. Since this is an unpolarized measurement, and, hence,
independent of the nucleon spins, the spin-flavor condition of
eq.~(\ref{eq2}) implies $d\sigma/d\Omega$ must be an isoscalar,
meaning that np and pp scattering should be equal up to $1/N_c$
corrections. In fig. 1(a) we show $d\sigma/d\Omega$ versus the
scattering angle as measured in the nucleon's center of mass
frame for a proton beam with kinetic energy of $T_{lab}=$279 MeV
(which is just below the pion production threshold) using data
from ref.~\cite{NNOnline}. For the reason mentioned above, we have
omitted forward angles.

\begin{figure}
\includegraphics[0in,0in][3.25in,3.8in]{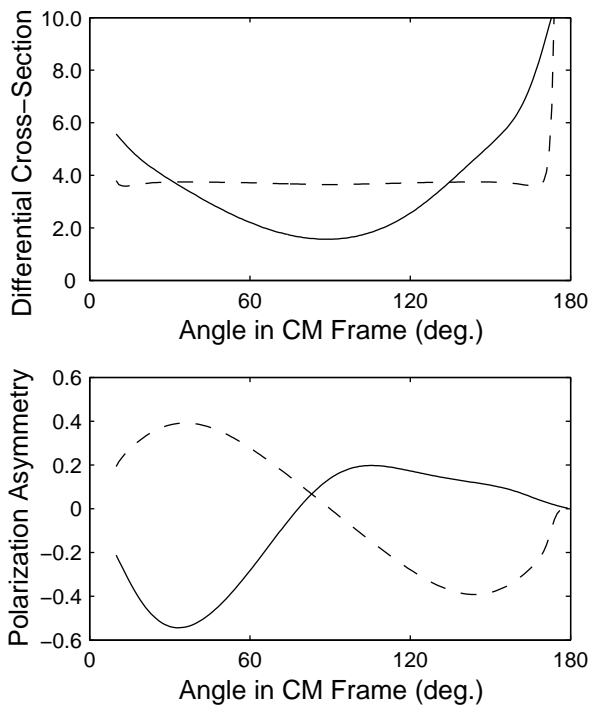}
\caption{(a) Elastic $d\sigma/d\Omega$ and (b) ${\cal A}_{pp}$
and $-{\cal A}_{np}$ for $T_{lab}=$~279 MeV as taken from
\cite{NNOnline}. The dashed line is for pp and the solid line is
for np.} \label{fig1}
\end{figure}

The polarization asymmetry, $\cal A$, quantifies the dependence
of  the differential cross section on the beam polarization.
Consider a target of protons (p) with a set spin ($m_t=+$, for
example) and a beam of nucleons N (either n or p) with set spins
($m_b=\pm$). The differential cross section obtained for this
arrangement will be written as $d\sigma_{m_b m_t}^N$ with the
$d\Omega$ suppressed since it is irrelevant here. The asymmetry
for pp (${\cal A}_{pp}$) and np (${\cal A}_{np}$) scattering is
then written as
\begin{eqnarray}
{\cal A}_{pp}=\frac{d\sigma_{+\:+}^p-d\sigma_{-\:+}^p}{d\sigma_{+\:+}^p+d\sigma_{-\:+}^p}\\
{\cal
A}_{np}=\frac{d\sigma_{+\:+}^n-d\sigma_{-\:+}^n}{d\sigma_{+\:+}^n+d\sigma_{-\:+}^n}.
\end{eqnarray}
Under spin and isospin reversal of the beam nucleons,
$d\sigma_{+\:+}^n\rightarrow d\sigma_{-\:+}^p$ and
$d\sigma_{-\:+}^n\rightarrow d\sigma_{+\:+}^p$, meaning that
eq.~(\ref{eq2}) implies that at leading order in $1/N_c$, ${\cal
A}_{np}=-{\cal A}_{pp}$. In figure 1(b) we show the $\cal A$
versus the scattering angle as measured in the nucleon's center
of mass frame for a proton beam with kinetic energy of
$T_{lab}=$~279 MeV.

Clearly, the large $N_c$ predictions for the two observables
considered above fail quite badly. These two cases are typical of
the disagreement found between the proposed spin-isospin
dependence and data collected from nucleon-nucleon elastic
scattering experiments in this kinematic range.  Moreover, we
cannot go to higher energies above 279 MeV because we will hit the
pion production threshold. Above the threshold, the elastic
differential cross section is not equal to the full collective
flow of baryon as baryons can flow in a given direction
accompanied by meson production. Thus, above threshold the elastic
differential cross section is not computable in TDMFT.

Why do the large $N_c$ predictions for these observables fail so
badly? One very plausible explanation for failure of the large
$N_c$ predictions with these kinematics is that the velocity is
insufficient to bring us to the required regime. As argued
before, the beam energy must be high enough to achieve the
semi-classical scattering regime. However, as the beam's kinetic
energy increases, it reaches the threshold for meson production.
Kinematically the threshold occurs at $T_{lab} = \frac{1}{2} M_N
v_{\rm t}^2 = 2 m_{\pi}$. Thus the maximum velocity before the
meson production channel opens is $v_{\rm t} = 2 \sqrt{m_\pi/M_n}
\sim N_c^{-1/2}$.  As argued above, the spin-flavor arguments here
depend on the interaction times being short compared to the
characteristic rotation time scales.  Were this not true,
additional terms proportional to $\vec{S}/{\cal I}$ and
$\vec{I}/{\cal I}$ could be added to eq.~(\ref{eq2}). In the
kinematics of ref.~\cite{TB} with $v\sim N_c^0$ the ratio of the
interaction time to the rotation time was of order $1/N_c$
implying that corrections due to these possible extra terms were
suppressed by $1/N_c$.  With the present kinematics the ratio is
$N_c^{-1/2}$ and the additional terms have a characteristically
smaller suppression. Of course, if $N_c$ is large enough one
expects the leading order expression to become increasingly
good.  However, for $N_c=3$ it is by no means clear that what
amounts to an expansion in $N_c^{-1/2}$ will usefully converge.

There is an additional issue which limits the usefulness of the
large $N_c$ expansion for these observables. The threshold
velocity for meson production is proportional to the square root
of the meson mass. Moreover, elastic scattering is only
associated with collective baryon flow and thus calculable by
TDMFT when one is below the threshold for {\it all mesons}. The
pion is the lightest meson and thus its threshold velocity sets
the maximum for which the elastic differential cross section is
computable in TDMFT. It is noteworthy that the pion mass is very
small on hadronic scales. This is due to the pseudo-Goldstone
nature of the pion and the small values of the up and down quark
masses. Thus, the pion mass is small for reasons which have
nothing to do with $N_c$. If we denote the ratio of $m_\pi$ to a
typical hadronic scale as $\beta$, the threshold velocity
$v_{\rm t} = 2 \sqrt{m_\pi/M_n} \sim \beta^{1/2} N_c^{-1/2}$.
Therefore, chiral effects suppress the velocity threshold by a
factor of $\beta^{1/2}$ relative to the natural scales in $1/N_c$
counting. This suppression could be responsible for pushing the
elastic regime below the regime for which the large $N_c$
expansion is useful.

In summary, we have shown in principle that the spin and flavor
dependence of the elastic differential cross section in
non-forward directions at energies just below the meson production
threshold can be computed in large $N_c$ QCD.  In practice, the
empirical scattering data is poorly described by the large $N_c$
relations.  We attribute this to the slow convergence of the
expansion which is, in effect, an expansion in $(\beta
N_c)^{-1/2}$, where $\beta$ is the ratio of $m_\pi$ to typical
hadronic scales.

The support of the U.S. Department of Energy for this research
under grant DE-FG02-93ER-40762 is gratefully acknowledged.

\end{document}